**THE EUROPEAN
PHYSICAL JOURNAL A**

Review

# The $^{22}$Ne($\alpha$,n)$^{25}$Mg reaction - state of the art, astrophysics, and perspectives

**Andreas Best**[1,2,a], **Philip Adsley**[3,4], **Ryan Amberger**[3,4], **Umberto Battino**[1,5], **Thomas Chillery**[6], **Marco La Cognata**[7], **Richard James deBoer**[8], **Daniela Mercogliano**[1,2], **Shuya Ota**[9], **David Rapagnani**[1,2], **Ragandeep Singh Sidhu**[10], **Roberta Spartà**[7,11], **Aurora Tumino**[7,11], **Michael Wiescher**[8]

[1] Department of Physics, University of Naples Federico II, Via Cintia, Naples 80126, NA, Italy
[2] Sezione di Napoli, INFN, Via Cintia, Naples 80126, NA, Italy
[3] Department of Physics & Astronomy, Texas A&M University, 578 University Drive, College Station 77843-4242, USA
[4] Cyclotron Institute, Texas A&M University, College Station 77843-3366, USA
[5] School of Chemical and Physical Sciences, Keele University, Lennard–Jones Laboratories, Keele ST5 5BG, Staffordshire, UK
[6] Laboratori Nazionali del Gran Sasso, Istituto Nazionale di Fisica Nucleare, Via G. Acitelli 22, L'Aquila - Localita Assergi 67100, Italy
[7] Laboratori Nazionali del Sud, Istituto Nazionale di Fisica Nucleare, Via S. Sofia 62, Catania 95100, Italy
[8] Department of Physics and Astronomy, University of Notre Dame, Notre Dame, IN 46556, USA
[9] National Nuclear Data Center, Brookhaven National Laboratory, Building 817, Upton 11973-5000, USA
[10] School of Mathematics and Physics, University of Surrey, Guildford GU2 7XH, UK
[11] Dipartimento di Ingegneria ed Architettura, Università degli Studi di Enna "Kore", Cittadella universitaria, Enna 94100, Italy

**Abstract** One of the most important stellar neutron sources is the $^{22}$Ne($\alpha$, $n$)$^{25}$Mg reaction, which gets activated both during the helium intershell burning in asymptotic giant branch stars and in core helium and shell carbon burning in massive stars. The $^{22}$Ne($\alpha$, $n$)$^{25}$Mg reaction serves as the main neutron producer for the weak $s$-process and provides a short but strong neutron exposure during the helium flash phase of the main $s$-process, significantly affecting the abundances at the $s$-process branch points. The cross section needs to be known at very low energies, as close as possible to the neutron threshold at $E_\alpha = 562$ keV ($Q = -478$ keV), but both direct and indirect measurements have turned out to be very challenging, leading to significant uncertainties. Here we discuss the current status of the field, including recent and upcoming measurements, and provide a discussion on the astrophysical implications as well as an outlook into the near future.

## 1 Introduction

The reaction $^{22}$Ne($\alpha$, $n$)$^{25}$Mg ($Q = -478$ keV [1]) is of great importance in the nucleosynthesis of the heavy elements beyond iron. Together with $^{13}$C($\alpha$, $n$)$^{16}$O, it is a neutron source of the main $s$-process, responsible for the production of elements with $90 < A < 204$, taking place in the hydrogen-helium intershell layer of thermally pulsing asymptotic giant branch (AGB) stars. $^{22}$Ne($\alpha$, $n$)$^{25}$Mg is activated during the short period of the helium flash when the temperature has risen high enough ($\gtrsim 0.25$ GK) for generating sufficient energy to overcome the negative $Q$ value of the reaction. This causes a short but intense neutron exposure that significantly affects the nucleosynthesis of isotopes at the $s$-process branch points: $^{95}$Zr, $^{135}$Ba, $^{141}$Ce, $^{152}$Gd, $^{170}$Tm, and $^{176}$Lu [2].

In massive stars, $^{22}$Ne($\alpha$, $n$)$^{25}$Mg is the main source of neutrons for the weak $s$-process, which is responsible for the production of elements with $60 < A < 90$. It is activated via $\alpha$ capture on the $^{14}$N ashes of the CNO cycle that fueled the preceding hydrogen burning phase. This fuel material is rapidly converted by the reaction sequence $^{14}$N($\alpha$, $\gamma$)$^{18}$F($\beta^+\nu$)$^{18}$O($\alpha$, $\gamma$)$^{22}$Ne, during the end of core He burning (in the contraction phase of the helium core) and also during C burning, at temperatures of ca. 0.25 GK and 1 GK, respectively [3–8].

The reaction $^{22}$Ne($\alpha$, $\gamma$)$^{26}$Mg competes with the neutron channel and can be activated before the stellar environment has reached the necessary temperature to overcome the negative $Q$ value of the neutron source. Some astrophysical simulations have shown the importance of the $^{22}$Ne($\alpha$, $\gamma$)$^{26}$Mg reaction at low energies, as the reaction could consume $^{22}$Ne before the stellar temperature becomes high enough to activate the $(\alpha, n)$ reaction, which ultimately affects the final $s$-process abundances [9–11]. The exact temperature at which the neutron channel becomes dominant over the $\gamma$ one depends on the reaction rates of both reactions and hence on their low-energy cross sections [12,13]. In addition, the interplay between the two channels has a strong effect on the production in Type II supernovae of $^{60}$Fe and $^{26}$Al, two radioactive isotopes observed in the Universe [14]. Another important astrophysical observable is the ratio of the magnesium isotopes, one of the few which can be observed directly with telescopes [15–17] but are also found in presolar grains [18]. To reliably describe the production of all these different species, an improved knowledge of both reaction rates is required down to temperatures of 0.2 GK.

The measurement of the very low reaction cross sections at the corresponding energies $< 1$ MeV is very difficult and so far it has only been possible to directly directly access the $^{22}$Ne$(\alpha, n)^{25}$Mg cross section down to a strong resonance at $E_{\text{c.m.}} = 706$ keV ($E_\alpha = 834$ keV). From there down to the neutron threshold at $E_\alpha = 562$ keV only upper limits are known. This is mostly due to the high neutron background in surface laboratories, not lower than 100 neutrons/hour even with dedicated shielding, hampering the identification of a clear neutron signal within the background [19,20]. The neutron background is due to cosmogenic neutrons impinging on the detector, but can also be caused by radiogenic high energy $\alpha$ particles from the natural decay chains which cause $(\alpha, n)$ reactions in the environment surrounding the detector setup, or from radioactive contaminants in the detector itself [21,22]. The measurements are also handicapped by boron, carbon and possibly beryllium impurities in the target material, which may trigger the beam-induced background neutron sources $^9$Be$(\alpha, n)$, $^{10}$B$(\alpha, n)$, $^{11}$B$(\alpha, n)$, and $^{13}$C$(\alpha, n)$ [23–28],[1] as observed in Refs. [29,30].

In addition to direct measurements, the states of interest above the $\alpha$ separation energy $E_\alpha = 10614.74 \pm 0.03$ keV [1] in the compound nucleus $^{26}$Mg have been under intense investigation in order to indirectly infer the resonance strength and hence the cross section [10,31–37]. Recently, the possible impact of known states near the $\alpha$- and neutron thresholds on the stellar reaction rate and its uncertainties has been investigated by Adsley et al. and Wiescher et al., incorporating the new results obtained since a comprehensive overview by Longland et al. in 2012 [11,13,38]. The possible contribution of $\alpha$-cluster states below the lowest directly measured $E_{\text{c.m.}} = 706$ keV resonance relies on the determination of spin-parities and partial widths and still comes with large uncertainties. The 706 keV resonance itself is thought to dominate at the 0.3 GK astrophysical scenarios but even for this relatively well-investigated resonance remain open question regarding its spin-parity, exact strength and of its identity with an observed resonance in the competing $(\alpha, \gamma)$ channel.

This paper is the result of a series of discussions that culminated in a workshop at the University of Naples "Federico II" in the spring of 2024 and a satellite meeting at the ChETEC general assembly in Strasbourg shortly thereafter. We aim to synthesize the main outcomes, incorporating the newest results since the publication of the last review, and to give an outlook into the near future, highlighting both planned direct and indirect measurements.

## 2 State of the art

### 2.1 Recent direct results

The $E_{\text{c.m.}} = 706$ keV resonance, dominant at astrophysical temperatures above 0.25 GK, was intensively studied throughout 1989–2001 through a series of direct measurements (Fig. 1). Table 1 summarises the reported resonance strengths. A first experiment [41] used an extended windowless gas target system aimed at the study of the $^{22}$Ne$(\alpha, \gamma)^{26}$Mg reaction. The study also provided information about the neutron channel based on the analysis of secondary $\gamma$-ray decays from neutron populated states and neutrons measured directly in $^3$He ionisation chambers. While this method was successful at measuring the cross sections of higher energy resonances, the efficiency of the detectors was insufficient to observe the resonance at 706 keV. Four subsequent studies [19,29,42,43] used the Stuttgart windowless gas target RHINOCEROS with slightly differing detector configurations, always using $^3$He counters, to successfully measure the $(\alpha, n)$ strength of the 706 keV resonance for the first time. However, the reported strengths spanned a rather large range between 80 and 180 $\mu$eV, almost all of them agreeing within their uncertainties with the exception of Ref. [43]. Another direct study [44] was performed using the 1 MV JN accelerator at the University of Toronto. There solid implanted $^{22}$Ne targets were used instead of a gas target, with neutrons again measured using $^3$He counters. This study reported a significantly larger resonance strength of $234 \pm 77$ $\mu$eV, a factor $\approx 2–3$ times larger than the strengths measured in Stuttgart. The authors reported a strong beam-induced background from the $^{13}$C$(\alpha, n)^{16}$O reaction which dominates their neutron yield above $E_{\text{lab}}^\alpha \sim 650$ keV. This background resulted from carbon deposition on the target and collimators, and was investigated and subtracted via measurements with a $^{20}$Ne target. The non-trivial beam-induced background may explain the discrepancy between this Toronto measurement and the Stuttgart measurements.

---

[1] The reaction $Q$ values are, in increasing order of target mass: 5701.0 $\pm$ 0.1 keV, 1058.7 $\pm$ 0.3 keV, 157.88911 $\pm$ 0.00022 keV and 2215.6093 $\pm$ 0.0003 keV [1].

**Fig. 1** Level diagram for the $^{22}$Ne$(\alpha, n)^{25}$Mg and $^{22}$Ne$(\alpha, \gamma)^{26}$Mg reactions. Excitation energies ($E_x$) in MeV, $J^\pi$ assignments from NNDC [39], with exception of $E_x = 11.153$ MeV state assignment taken from [31,40]

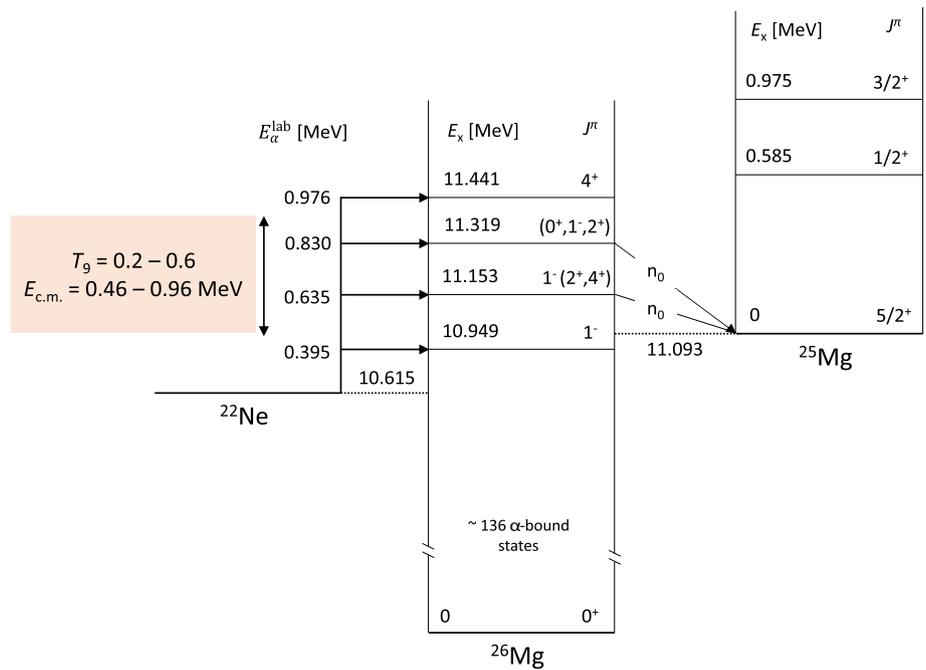

**Fig. 2** Comparison of $\omega\gamma_{(\alpha,n)}$ for the $E_{c.m.} = 706$ keV resonance. Adapted from Figure 7 of Shahina et al. [30]

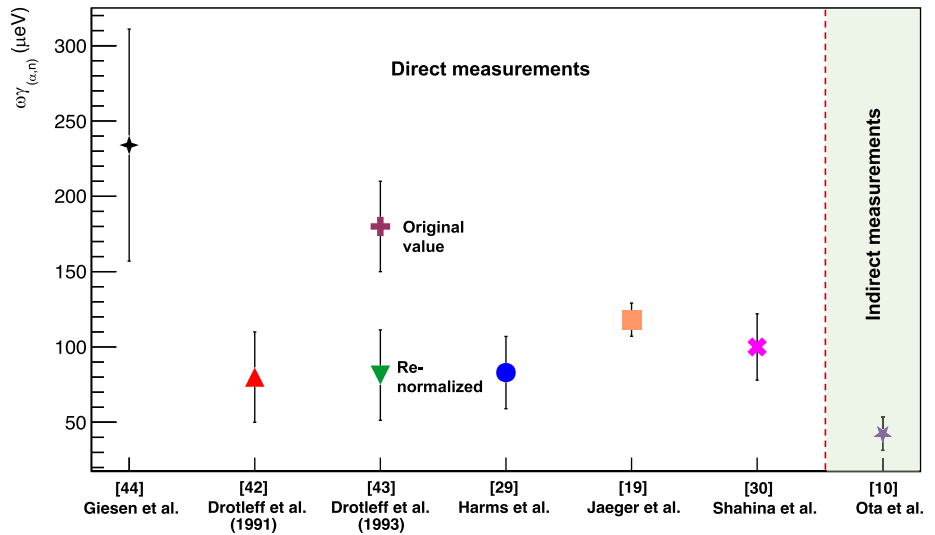

For over 20 years, the $E_{c.m.} = 706$ keV resonance strength remained untouched by direct measurements until the very recent study reported in 2024 by Shahina et al. [30]. This recent measurement was performed at the University of Notre Dame Nuclear Science Laboratory using the St. ANA Pelletron accelerator. An $\alpha$-particle beam of energy 910 keV was used to bombard a solid target consisting of $^{22}$Ne ions ($6.21(37) \times 10^{17}$ atoms/cm$^2$) implanted on 0.5 mm thick Ta backings. A total charge of $\approx 3$ C was deposited on the target, with the beam energy selected to reach the thick-target yield plateau of the 706 keV resonance. Neutrons from beam-target interactions were measured using a stilbene organic scintillator mounted directly behind the target ($\theta = 0°$) in close geometry, with an efficiency of 5.81(25)%[2] for detection of the 285 keV reaction neutrons. The extracted strength from this study, 100(22) μeV, is in good agreement with the Stuttgart measurements [19,29,42,43].

A summary of the measured resonance strengths from Ref. [30] is adapted here in Fig. 2. For this, a renormalization of the 706 keV resonance strength reported in the 1993 work[43] by Drotleff et al. was performed through comparison of the relative strength of the $E_r = 1337$ keV resonance with that reported by the Stuttgart group, reducing the 706 keV strength by a factor of two. Unlike for the 706 keV resonance, the 1337 keV resonance is comparable in width to the energy loss in the employed gas target, making a direct scaling difficult, and the use of other more narrow resonances in the high energy region only would lead to a scaling of the order of ∼ 15% [45], leading to a quite large uncertainty in this scaled value [46].

---

[2] Efficiency from [30] taken as weighted average of 5.88(34)% and 5.72(37)% from $^7$Li(p,n)$^7$Be and $^{51}$V(p,n)$^{51}$Cr, respectively.

**Fig. 3** State-of-art of $^{22}$Ne($\alpha$, $n$)$^{25}$Mg reaction yield measurements reported in the literature. At the tails of resonances, the yields can be taken as cross section in $\mu$b. The red bar designates the lowest energy range of interest. Adapted from Jaeger et al. [19] Figure 1

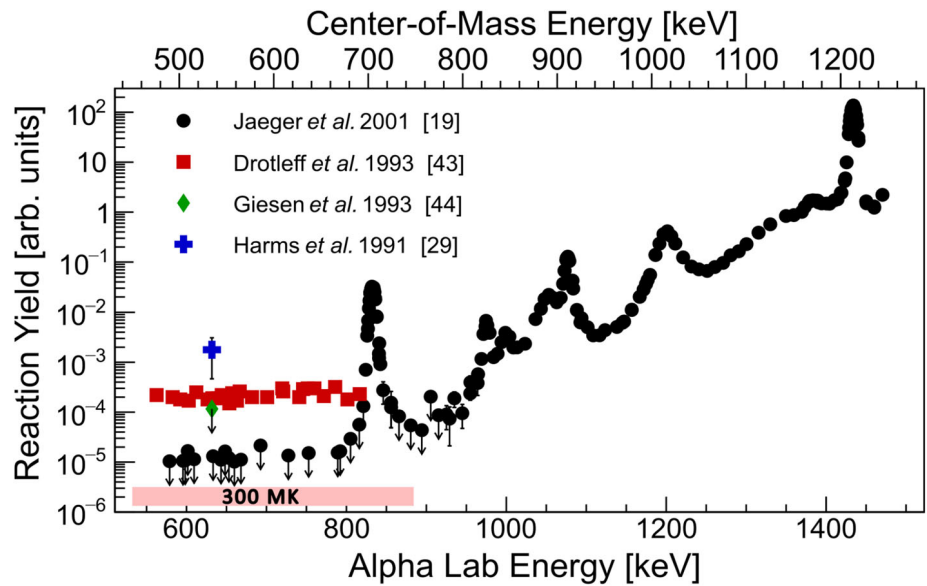

**Table 1** Nean reaction $E_{c.m.}$ = 706 keV and 540 keV resonance strengths from the literature

| Reference | $\omega\gamma_{(\alpha,n)}$ [$\mu$eV] | |
|---|---|---|
| | 706 keV | 540 keV |
| Wolke et al. [41] | $\leq 340$ | – |
| Harms et al. [29] | $83 \pm 24$ | – |
| Drotleff et al. [42] | $80 \pm 30$ | – |
| Drotleff et al. [43] | $180 \pm 30$ | – |
| [a]Drotleff et al. [43] | $81 \pm 30$ | – |
| Giesen et al. [44] | $234 \pm 77$ | $\leq 5$ |
| Jaeger et al. [19] | $118 \pm 11$ | $\leq 0.06$ |
| Shahina et al. [30] | $100 \pm 22$ | – |
| [b]Ota et al. [10] | $42 \pm 11$ | – |

[a] Renormalised. [b] Indirect measurement

Figure 3 shows the present state-of-art of the low-energy resonances ($E^{\alpha}_{lab} < 1.5$ MeV) as a function of incident $\alpha$-particle energy. The latest measurement of Jaeger et al. [19] in 2001 remains the most sensitive as indicated in Fig. 3 and Table 1. This measurement clearly covers the 706 keV resonance that was discussed previously. The work of Jaeger et al. alongside other historical measurements [19,29,44] extend further down to $E_{c.m.}$ ∼ 450 keV covering across the astrophysical energy region for AGB stars (∼ 0.3 GK). However these reported results represent upper limits to the yield, leaving the ($\alpha$, $n$) rate weakly constrained for these environments.

Besides the relatively strong $E_{c.m.}$ = 706 keV resonance discussed so far, a weak resonance at $E_{c.m.}$ ≈ 540 keV is posited to exist, with thus far only upper limits quoted on the resonance strength, see Table 1. The first direct observation of this resonance is tentatively reported by Fig. 7 of Harms et al. [29], however resonance properties were unable to be determined due to limited covered energies and statistics. The low-energy region between $E_{c.m.}$ = 450–700 keV was covered by Drotleff *et al.* [43], however no discernible resonance structure was observed despite a factor ≈ 10 improvement in sensitivity over the Harms study. The situation was improved by Giesen et al. [44] who, despite strong beam-induced background from $^{11}$B($\alpha$, $n$)$^{14}$N, extracted an upper limit of ≤ 5 $\mu$eV for the $E_{c.m.}$ ≈ 540 keV. The latest, and most sensitive, direct measurement covering this low-energy resonance remains Jaeger et al. [19], quoting an upper limit of < 60 neV.

A direct measurement of the $^{22}$Ne($\alpha$, $\gamma$)$^{26}$Mg channel has been performed at the Laboratory for Experimental Nuclear Astrophysics (LENA), at Chapel Hill, North Carolina [47]. A single charged He beam was delivered by the 1-MV JN Van de Graaff accelerator on a blister-resistant target, fabricated by implanting $^{22}$Ne ions into a porous titanium structure. To further improve the signal-to-background ratio, a $\gamma\gamma$-coincidence spectrometer was employed. This setup consists of a coaxial high-purity germanium with 135% relative efficiency, located at 1.1 cm from the target at 0° relative to the beam direction. In addition, a NaI(Tl) annulus and a plastic ("veto") scintillator shield surrounded the HPGe, reducing the $\gamma$-ray background below 2.6 MeV by several orders of magnitude. The data acquired on resonance and off resonance resulted in a new estimation of the resonance energy $E_{c.m.}$ = 706.6 ± 2.5 keV and resonance strength $\omega\gamma$ = (46 ± 12) $\mu$eV. In addition, two transitions corresponding to $E_x \rightarrow$ 1809 keV and $E_x \rightarrow$ 7062 keV were observed and their branching were estimated to be 54 ± 12 % and 64 ± 12 %, respectively. A second measurement of this resonance was done recently at the University of Notre Dame by Shahina et al. [40], using implanted $^{22}$Ne targets and a summing NaI detector. No determination of the resonance energy was done, but the strength was determined as $\omega\gamma_{(\alpha,\gamma)}$ = (35 ± 4) $\mu$eV, in agreement with the Hunt et al. result. This Notre Dame study also attempted to directly measure the $E_{c.m.} \approx$ 540 keV resonance strength for the first time. They report an upper limit of $\omega\gamma_{(\alpha,\gamma)} < 0.15$ $\mu$eV, with a tentative spin-parity assignment 4$^+$ of the corresponding $^{26}$Mg $E_x$ = 11.153 MeV state.

The higher energy region above the $E_{c.m.}$ = 706 keV resonance has been comprehensively measured only by the Stuttgart group [19,41], see Fig. 3 for the neutron channel. Directly above the low-energy resonance, there is a region of upper limits, and it is possible that there are some unresolved states further towards higher energies. While this energy region is not considered crucial for astrophysics, to obtain a reliable reaction rate, one requires an extrapolation of the resonance flanks into the low-energy region, usually done by an $R$-matrix formalism. For this and to cross-check the previous results, a re-measurement of the high-energy region would also be beneficial.

## 2.2 Indirect results

Indirect measurements have played important roles in $^{22}$Ne $(\alpha, n/\gamma)$ reactions as they can constrain the strength of resonances ($\omega\gamma$) at low energy where the cross sections are too small for accurately determining the resonance strengths by direct measurements [11,13,38]. Unlike direct measurements, in which the $\omega\gamma$ is directly obtained from the excitation function of the cross sections, individual resonance parameters such as $E_x$, $J^\pi$, and $\Gamma_{n/\gamma/\alpha}$ need to be deduced in the indirect measurements. For example, transfer, inelastic scattering, neutron-capture and transmission, and photon scattering/photo-neutron measurements, including information on decay pathways, have been successfully used to constrain these parameters of the low-energy resonances [38]. These techniques typically cannot determine all the resonance parameters at once, and those deduced by different techniques need to be put together to infer $\omega\gamma$. Nevertheless, indirect measurements are currently the only means for identifying some resonances that can contribute to the $^{22}$Ne+$\alpha$ reaction, in particular at the lower energies so far inaccessible to direct measurements.

As such, establishing indirect techniques that allow extraction of reliable values of the resonance parameters is extremely important for the study of $^{22}$Ne+$\alpha$ reaction rates. As stated in the introduction, even the $\omega\gamma$ for the most impor-tant resonance, namely $\omega\gamma_{(\alpha,n)}$ and $\omega\gamma_{(\alpha,\gamma)}$ at $E_{c.m.}$ = 706 keV, have not been constrained well enough yet, despite being clearly observed in multiple direct measurements. Discrepancies between different direct measurements have been reported for the resonances at higher energy as well in Table II in Ref. [19].

As another example for the utility of indirect measurements, the $E_{c.m.}$ = 334 keV ($E_x$ = 10.95 MeV) resonance lies below the neutron threshold ($S_n$) and therefore nearly exclusively proceeds through the $(\alpha, \gamma)$ reaction. While the current upper limit of its $\omega\gamma$ achieved by a direct measurement is $< 10^{-10}$ eV [48], the different indirect measurements reach higher sensitivities and consistently indicate $\omega\gamma \lesssim 10^{-13}$ eV [31,49,50]. Similarly, the $(\alpha, \gamma)$ strength for the possible $E_{c.m.}$ = 557 keV resonance was recently constrained to be $< 150$ neV [40], while $\lesssim 1$ neV is reported by a different indirect measurement [33] (see their Table 1 and the possible $\Gamma_n / \Gamma_\gamma$ in Table 3 of Ref. [32]).

Among the different indirect techniques, one effective approach is using $\alpha$-transfer reactions due to their selectivity for $\alpha$-cluster states. Because the $J^\pi$ of both $^{22}$Ne and $\alpha$ ground states is 0$^+$, only natural parity states with high $\alpha$-spectroscopic factors are of primary interest for both the $(\alpha, \gamma)$ and $(\alpha, n)$ reaction channels. The selectivity for $\alpha$-induced resonances achieved by $\alpha$ transfer reactions is paramount given the high level-density of $^{26}$Mg compound nucleus. ($^6$Li,$d$) and ($^7$Li,$t$) reactions are almost exclusively used for that purpose [31–33,44]. Other techniques using neutron beams at n-TOF or elsewhere [36,51,52], $(\gamma, \gamma/\gamma')$ [49,53], $(\gamma, n)$ [54,55], and $(d, p)$ [37], $(\alpha, \alpha')$ [31,34] and $(p, p')$ [35] measurements with magnetic spectrographs allowing determination of the excitation energies of the populated states with more than an order-of-magnitude enhanced energy resolution. However, unlike the $\alpha$-transfer reactions, these reactions populate many irrelevant resonances for the $^{22}$Ne+$\alpha$ reactions and struggle to clearly identify the resonances of importance. In the $\alpha$-transfer reactions, the information on the $J^\pi$ and $\Gamma_\alpha$ (or $\alpha$-spectroscopic factors) can be inferred by measuring the angular distribution of the light ejecta ($d$ or $t$) and comparing it with DWBA calculations (see e.g., Refs. [31,32,44]). Since the reaction mechanism and optical potentials needed for the DWBA calculations are not as reliably known as single-particle transfer reactions such as $(d, p)$, care must be taken to unambiguously extract $J^\pi$ and $\Gamma_\alpha$ e.g., by experimentally constraining the optical potentials. Independent experiments to constrain the optical potentials are thus also highly welcome.

As $\Gamma_\alpha$ is much smaller than $\Gamma_n$ and $\Gamma_\gamma$ for resonances of our interest, $\omega\gamma \simeq (2J+1)\Gamma_\alpha$ [11]. The $(\alpha, \gamma)$ and $(\alpha, n)$ strengths are then determined by multiplying the $\omega\gamma$ with the branching ratio, $\Gamma_\gamma/\Gamma$ or $\Gamma_n/\Gamma$, where $\Gamma$ is the total width (= $\Gamma_\alpha + \Gamma_n + \Gamma_\gamma$; note $\Gamma_n$ = 0 for resonances below $S_n$). Neutron spectroscopic factors, which can be studied well by ($d$, $p$) reactions, also could play an important role in deducing $\Gamma_n/\Gamma_\gamma$ ratios that depend on the single-particle component in the resonance state [11].

Recent $\alpha$ transfer studies performed at Texas A&M University (TAMU) directly measured the $\Gamma_n/\Gamma_\gamma$ ratio by comparing the $^{25}$Mg and $^{26}$Mg events from the $E_{\text{c.m.}} = 706$ keV resonance populated by the ($^6$Li,$d$) reaction in inverse kinematics [10,32]. This approach is effective in particular for the 706-keV resonance as the ($\alpha$, $\gamma$) and ($\alpha$, $n$) strengths compete in the same order of magnitude ($\Gamma_n/\Gamma_\gamma = 1–7$). The approach is also independent of theoretical uncertainties such as optical potentials. While a broad $E_x$ range of $^{26}$Mg states were observed in the experiment, no $\Gamma_n/\Gamma_\gamma$ was reported for the possible $E_{\text{c.m.}} = 557$ keV resonance due to the limited energy resolution. Nevertheless, as no significant counts were observed in the ($\alpha$, $n$) channel ($^{25}$Mg gated spectrum) at $E_{\text{c.m.}} \lesssim 557$ keV, a rather weak $\omega\gamma_{(\alpha,n)}$ were inferred for low energy resonances ($E_{\text{c.m.}} = 557$ keV and below). For instance, $\Gamma_\alpha < 0.22$ neV was deduced for the $E_{\text{c.m.}} = 498$ keV ($E_x = 11.11$ MeV) resonance, which is quite smaller than 2.7 neV adopted as an upper limit in Ref. [36] (see also discussion in Ref. [10]).

Another $\alpha$-transfer experiment performed at TAMU measured the Asymptotic Normalization Coefficients (ANCs) for four resonances ($E_{\text{c.m.}} = 207$, 334, 466, and 706 keV; corresponding to $E_x = 10.83$, 10.95, 11.08, and 11.32 MeV, respectively) by ($^6$Li,$d$) and ($^7$Li,$t$) reactions at sub-Coulomb energies [33]. Unlike the $\alpha$-spectroscopic factors, the ANCs more directly represent the amplitude of the $^{22}$Ne-$\alpha$ cluster state wave function at the exterior of the nuclear potential. Thus, the magnitude of the ANCs is an effective way to probe the $\Gamma_\alpha$, and consequently $^{22}$Ne+$\alpha$ rate at stellar energy. Due to the sub-Coulomb barrier energy, the $\alpha$-transfer reactions are peripheral and dominated by Coulomb interaction. Since the nature of Coulomb potential is much better known than nuclear potentials, the ANC and $\omega\gamma$ inferred from the experiment is less model dependent. The $\alpha$-particle partial width inferred for the $E_{\text{c.m.}} = 706$ keV resonance (assuming $J^\pi = 0^+$, $\Gamma_\alpha = 61 \pm 4_{\text{(stat)}} \pm 10_{\text{(syst)}}$ μeV) is consistent with the $\Gamma_\alpha = 79(13)$ μeV deduced by the other TAMU experiment [32]. Note that, since the sub-Coulomb transfer reaction only probes the $\alpha$-particle partial width, the comparison made should be between the inferred $\Gamma_\alpha$ in contrast to the $\omega\gamma$ for each individual channel which requires some assumption to be made for the relative $\gamma$-ray and neutron branching ratios. The $E_{\text{c.m.}} = 557$ keV resonance was not observed ($\omega\gamma < 65$ peV), indicating that the contribution of the resonance to the stellar reaction rates is possibly negligible. The results from the two TAMU experiments are summarized in Table 2.

**Table 2** $\Gamma_\alpha$ obtained from the two recent TAMU experiments (for adopted $J^\pi$; $0^+$ (706 keV), $2^+$, $2^+$, $2^+$, $1^-$, and $2^+$ (207 keV)). Both works adopted the same $J^\pi$. For the results from Jayatissa et al., the first uncertainties are statistical and the others are systematic

| $E_{\text{c.m.}}$ | $\Gamma_\alpha$ [eV] | |
|---|---|---|
| | Jayatissa et al. [33] | Ota et al. [32] |
| 706 | $(6.1 \pm 0.4 \pm 1.0) \times 10^{-5}$ | $(7.9 \pm 1.3) \times 10^{-5}$ |
| 557 | $< 1.3 \times 10^{-11}$ | Adopted from [33] |
| 498 | – | $< 2.2 \times 10^{-10}$ |
| 466 | $(5.7 \pm 0.7 + 1.4 - 1.2) \times 10^{-11}$ | Adopted from [33] |
| 334 | $(3.0 \pm 0.3 + 0.8 - 0.6) \times 10^{-14}$ | Adopted from [33] |
| 207 | $(2.1 \pm 0.3 \pm 0.4) \times 10^{-22}$ | Adopted from [33] |

## 3 Astrophysics impact

Both the $^{22}$Ne($\alpha$, $n$)$^{25}$Mg and $^{22}$Ne($\alpha$, $\gamma$)$^{26}$Mg reactions significantly impact the neutron-capture nucleosynthesis of elements heavier than iron via the $s$-process in both low- and intermediate-mass (M $< 8$ M$_\odot$) and high-mass stars (M $> 8$ M$_\odot$) [7,8,56,57], hosting the main and weak components of the $s$-process, respectively. The main $s$-process is responsible for producing approximately half of the elements between Zr and Pb observed in the Solar System, while the weak $s$-process dominates between Fe and Zr.

In recent years, the reaction rates of both $^{22}$Ne($\alpha$, $n$)$^{25}$Mg and $^{22}$Ne($\alpha$, $\gamma$)$^{26}$Mg have been re-evaluated by different research groups, leading to contradictory results. For example, Ref. [11] showed that their newly evaluated $^{22}$Ne($\alpha$, $n$)$^{25}$Mg rate is a factor of three higher at typical He-burning temperatures (0.2 GK $< T < 0.3$ GK) and a factor of two lower at typical C-burning temperatures $T \sim 1$ GK than the rates provided in Ref. [38] including the TAMU results from Refs. [32,33].

A comparison of some of the most widely used reaction rates for stellar nucleosynthesis simulations is presented in Fig. 4. Angulo et al. [58] calculated the upper limit of the rate (for T $< 0.4$ GK) by including all known states with strengths according to the Wigner limit, hence using the maximum physically allowed values. In contrast, Ref. [19] and later works added only one additional resonance at 635 keV to the list of those used for their recommended rates, and then used the upper limit of only the already included resonances in order to calculate their "high" reaction rate. At He-burning temperatures ($T \sim 0.3$ GK), the reaction rate including the TAMU results of Refs. [32,33] from Ref. [38] is approximately a factor of three lower than other rates due to their inclusion of a substantially lower ($\alpha$, $n$) resonance strength for the $E_r = 706$ keV resonance. Conversely, at C-burning temperatures ($T \sim 1$ GK),

**Fig. 4** The $^{22}$Ne$(\alpha, n)^{25}$Mg reaction rates from Refs. [19,38,58], and [11], relative to the median value of Ref. [13], are shown. The median value of each study is indicated by the line style in the legend, while the corresponding uncertainty band is represented by the shaded region of the same color. The rate from Ref. [38] includes the results from Refs. [32,33]

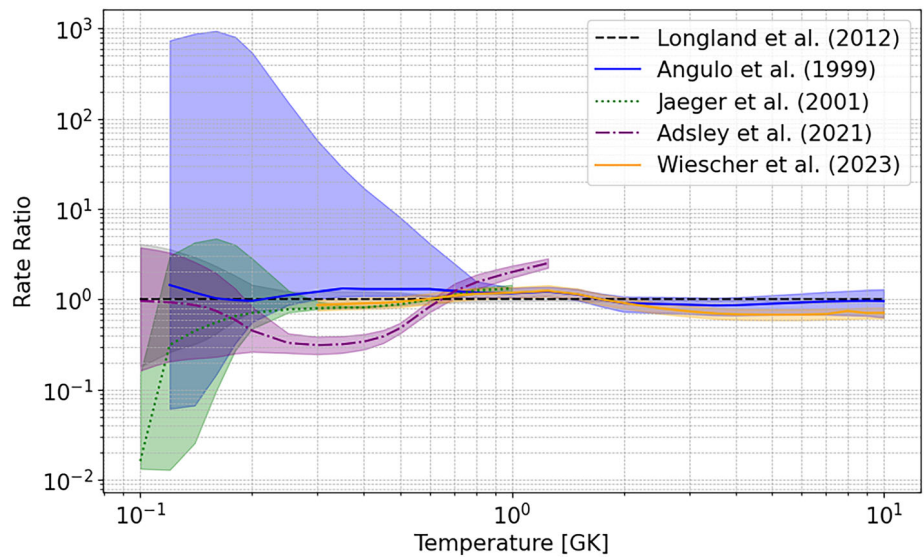

Ref. [38] reports a reaction rate higher by a factor of two compared to other studies, owing to the inclusion of a larger number of energy levels. The ambiguity in the choice of included resonances lends strong support for more experimental studies of properties of excited states in the relevant energy region in order to being able to decide on their possible contribution to the reactions.

## 3.1 Massive stars

In massive stars, the main neutron source is the $^{22}$Ne$(\alpha, n)^{25}$Mg reaction. After the exhaustion of central H, convective core He-burning is ignited via the triple-$\alpha$ reaction at temperatures around $2 \times 10^8$ K. $^{22}$Ne is produced from the initial CNO isotopes, which are first converted into $^{14}$N during the core H-burning phase and then into $^{22}$Ne via $^{14}$N$(\alpha, \gamma)^{18}$F$(\beta^+, \nu)^{18}$O$(\alpha, \gamma)^{22}$Ne during core He-burning. When the central temperature reaches $\sim 2.5 \times 10^8$ K, the $^{22}$Ne$(\alpha, n)^{25}$Mg reaction is activated, generating a peak neutron density of $\sim 10^7$ cm$^{-3}$ [6,59,60].

At the end of core He-burning, $^{22}$Ne is not completely consumed, and $^{22}$Ne$(\alpha, n)^{25}$Mg is re-ignited during C-burning, first in the core and then in a shell. However, core C-burning nucleosynthesis products are unlikely to be ejected and contribute only marginally to the enrichment of the interstellar medium. At the higher C-burning temperatures $T \sim 1$ GK, the neutron density increases beyond that reached during core He-burning, typically reaching around $10^{12}$ cm$^{-3}$. Given the longer timescales over which the $^{22}$Ne$(\alpha, n)^{25}$Mg reaction is active during core He-burning ($\sim 10^5$ years) compared to shell C-burning ($\sim 10$ years), the bulk of the s-process production takes place during core He-burning.

In Fig. 5, we show the results of our stellar nucleosynthesis simulations at the end of the core He-burning and shell C-burning phases, obtained by post-processing the stellar structure of the 15 M$_\odot$, Z = 0.006 model by Ref. [8] (where Z denotes the initial metallicity of the star, i.e., the total abundance in mass fraction of all elements heavier than He). This choice is motivated by the fact that the weak s-process efficiency is maximized at Z = 0.006 (see Refs. [8,61]). Additionally, 15 M$_\odot$ initial mass models are widely adopted by different groups in the literature for s-process nucleosynthesis simulations (see Refs. [7,8,56,60]).

In particular, we computed the full nucleosynthesis adopting the $^{22}$Ne+$\alpha$ reaction rates from different sources, specifically the rate from Ref. [13], the rate from Ref. [38] including the TAMU results from Refs. [32,33], and Ref. [11]. The impact on the resulting s-process nucleosynthesis is clearly visible. In particular, the factor of 3 difference in the production factors of the first peak elements is consistent with the differences between the employed reaction rates at He-burning temperatures.

This uncertainty in the final isotopic abundances, stemming from uncertainties in the $^{22}$Ne$(\alpha, n)^{25}$Mg and $^{22}$Ne$(\alpha, \gamma)^{26}$Mg reaction rates, also affects galactic chemical evolution studies aimed at computing the contribution of specific stellar sources to the isotopic abundances observed today in the Solar System, particularly those elements that receive a substantial contribution from the weak s-process ($26 < Z < 40$). An example of this is given in Table 3, which shows the Kr isotopic abundance ratios obtained from our weak s-process calculations relative to $^{82}$Kr, the most abundant s-only Kr isotope, as it is shielded from any r-process contribution by $^{82}$Se. This allows a rough estimate of the s-process contribution to each stable Kr isotope by comparing its abundance ratio relative to $^{82}$Kr from a given nucleosynthesis simulation to the same ratio in the Sun. Table 3 shows that the s-process contri-bution to $^{83}$Kr and $^{84}$Kr does not strongly depend on the $^{22}$Ne($\alpha$, n)$^{25}$Mg and $^{22}$Ne($\alpha$, $\gamma$)$^{26}$Mg reaction rates. However, the $s$-process contribution to $^{82}$Kr can vary by about a factor of 2. Specifically, the largest contribution is obtained when the $^{22}$Ne($\alpha$, n)$^{25}$Mg and $^{22}$Ne($\alpha$, $\gamma$)$^{26}$Mg reaction rates from Ref. [13] and Ref. [11] are adopted. This is due to the stronger activation of the branching $^{85}$Kr point (half-life of 10.72 years) during core He-burning. This uncertainty, originating from the selection of nuclear reaction rates, significantly affects the determination of the relative contribution of the $s$- and $r$-processes to Kr isotopes, as well as other elements influenced by neutron-capture branching points.

**Fig. 5** Elemental abundance distributions in mass fraction (upper panel) and production factors relative to solar abundance (lower panel) at the end of core He-burning and shell C-burning in a M = 15 $M_\odot$, Z = 0.006 massive star model from [8]. Each distribution corresponds to a specific selection of $^{22}$Ne+$\alpha$ reaction rates and to the end of a specific burning phase

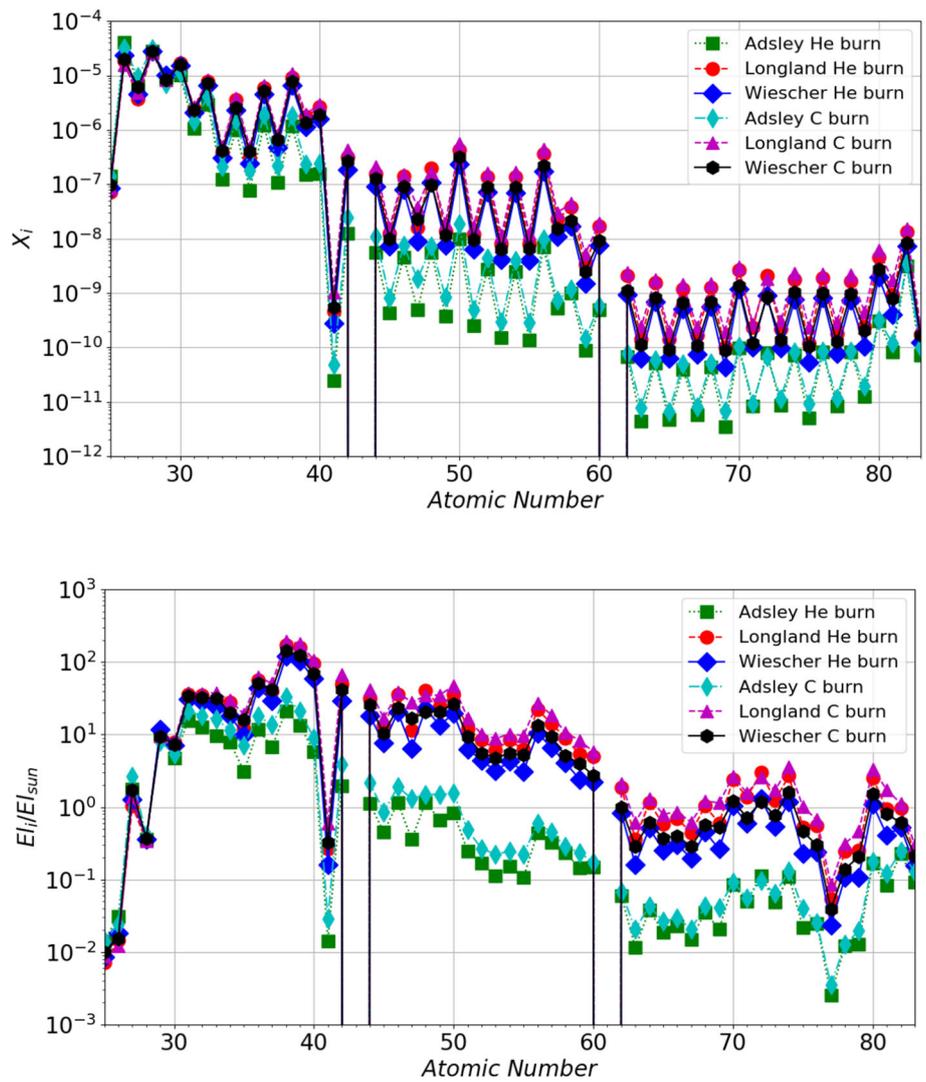

## 3.2 Low- and intermediate-mass stars

In low- and intermediate-mass stars, $^{22}$Ne($\alpha$, n)$^{25}$Mg is activated during the thermally pulsing asymptotic giant branch (TP-AGB) phase, i.e., the advanced evolutionary stage fol-lowing the red giant branch and preceding planetary nebula formation [57,62–64]. In particular, AGB stars experience recurrent He-flashes in the so-called He-intershell as a result of unstable He-burning in a shell configuration and the partial electron degeneracy of the plasma. During a He-flash, the temperature increases to ∼ 2.5 × 10$^8$ K, and a pulse-driven convective region develops, mixing the $^{14}N$-rich ashes of the H-burning layers above. This process produces $^{22}Ne$, which is then consumed via the $^{22}Ne(\alpha, n)^{25}Mg$ reaction. In this way, a peak neutron density of $\sim 10^{11}$ cm$^{-3}$ is produced, impacting several $s$-process isotopic ratios [65,66].

**Table 3** Abundance ratios of Kr isotopes relative to $^{82}$Kr, normalized to solar, at the end of shell C-burning. Theoretical abundances are from a M = 15 $M_\odot$, Z = 0.006 massive star model from [8] and have been computed using different $^{22}$Ne($\alpha$, n)$^{25}$Mg and $^{22}$Ne($\alpha$, $\gamma$)$^{26}$Mg reaction rates, as indicated in the column headers. The rate from Ref. [38] is that including the TAMU results from Refs. [32,33]

| Species (i) | (i/$^{82}$Kr)/(i/$^{82}$Kr)$_\odot$ | | |
|---|---|---|---|
| | (Adsley) | (Longland) | (Wiescher) |
| $^{83}$Kr | 0.49 | 0.50 | 0.50 |
| $^{84}$Kr | 0.43 | 0.51 | 0.50 |
| $^{86}$Kr | 0.11 | 0.21 | 0.20 |

**Fig. 6** Comparison of M = 3 $M_\odot$, Z = 0.03 and M = 2 $M_\odot$, Z = 0.02 stellar model results from [70] with measured Zr and Ba isotopic ratios from presolar SiC grains. The theoretical tracks show the surface abundances obtained using the $^{22}Ne(\alpha, n)^{25}Mg$ and $^{22}Ne(\alpha, \gamma)^{26}Mg$ reaction rates from different sources, as in Fig. 5. All the theoretical tracks start at the origin, i.e., from an initial solar value of the isotopic ratios considered. The "carbon star" phase is marked by the large symbols at the end of the tracks, each corresponding to a single third dredge-up episode during the carbon-rich phase. See main text for details

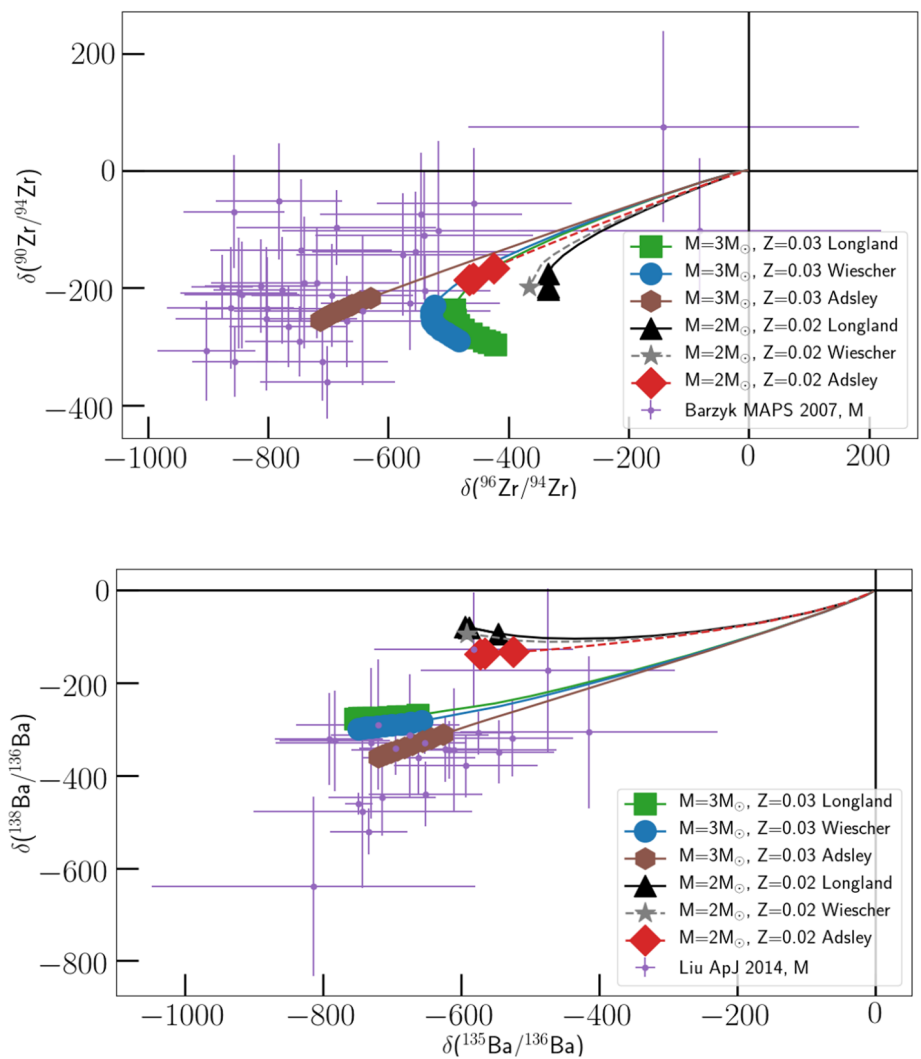

The energy released by the He-flash expands and cools the stellar envelope, increasing plasma opacity and the temperature gradient. As a result, convective efficiency increases, and material from the He-intershell is transported into the envelope and eventually reaches the stellar surface before being ejected and condensed into dust grains [67,68]. Some of these grains can be recovered from ancient meteorites, and their composition can be studied in the laboratory.

Since their discovery almost 40 years ago, these grains have provided powerful constraints for modeling nucleosynthesis in stars and related explosions. Moreover, the fingerprint of the nearly pure nucleosynthetic signatures carried by these grains can be exploited to understand the history of the early Solar System [68,69].

In Fig. 6, we compare the isotopic ratios reported in presolar silicon carbide (SiC) grains in Refs. [71] and [72] with those obtained from the surface composition of two AGB star models with initial M = 2, 3 $M_\odot$ and Z = 0.02, 0.03 in Ref. [70]. This choice is motivated by the results of Ref. [73], who found that presolar SiC grains in the Solar System predominantly originated from AGB stars with M $\sim$ 2 $M_\odot$ and Z $\sim$ Z$_\odot$ (consistent with Z = 0.02), while Refs. [67,68] proposed AGB stars with higher masses (M > 3$M_\odot$) and higher metallicities (Z = 0.03) as the parent stars of large (> 1 $\mu$m in diameter) mainstream grains.

Plotted values are given in $\delta$-value notation to represent the isotopic ratios, i.e., the permil variation with respect to the solar ratio (for which $\delta = 0$), so that $\delta = ((\text{model ratio/solar ratio}) - 1) \times 1000$.

All the theoretical tracks start at the origin, i.e., from an initial solar value of the isotopic ratios considered. As the star evolves, several mixing episodes enrich the surface with freshly synthesized products from the deep interior. As a con-sequence, the surface composition starts to diverge from the origin. When the star enters the AGB phase, third dredge-up episodes bring both $s$-process elements and C to the surface. Eventually, the surface C/O numerical ratio exceeds unity, which is the critical condition for SiC grains to condense in AGB stars' atmospheres. This "carbon star" phase is marked by the large symbols at the end of the tracks, each corresponding to a single third dredge-up episode during the carbon-rich phase.

Also in this case, the different $^{22}$Ne($\alpha, n$)$^{25}$Mg and $^{22}$Ne($\alpha, \gamma$)$^{26}$Mg reaction rates considered lead to different results and levels of agreement with laboratory measurements. In particular, the models employing the reaction rates in Ref. [38] including the experimental results from Refs. [32,33] seems to agree better with laboratory data, as the lower $^{22}$Ne($\alpha, n$)$^{25}$Mg rate leads to reduced neutron captures on the unstable $^{95}$Zr (hence producing less $^{96}$Zr) and to lower abundance accumulation on the neutron-magic isotope $^{138}$Ba.

However, these results also heavily depend on other highly uncertain stellar physics inputs, particularly the convective boundary mixing at the bottom of the intershell during the He-flash. A lower convective boundary mixing efficiency would result in a lower He-flash temperature and thus a lower $^{22}$Ne($\alpha, n$)$^{25}$Mg activation (see Refs. [70,74]), improving agreement with observations and reducing the need to favor a particularly low reaction rate. This highlights the urgency of reducing the large uncertainty affecting the $^{22}$Ne($\alpha, n$)$^{25}$Mg reaction rate to better distinguish its effects from those arising from stellar physics uncertainties.

# 4 Experimental outlook

Due to its still largely uncertain stellar reaction rate, $^{22}$Ne($\alpha, n$)$^{25}$Mg (and the competing channel $^{22}$Ne($\alpha, \gamma$)$^{26}$Mg) is under intense investigation. Indirect studies at the EMMA separator of TRIUMF, at TAMU and at INFN-LNS are in various stages from planning to preparation to already having data collected. By probing the compound nucleus near the $\alpha$ and neutron thresholds the aim is to resolve the conflicts between the above discussed experimental results. In addition, thanks to the expansion of deep underground laboratories across three continents, multiple direct measurements at low energies are upcoming or currently being conducted.[3] Here we give an overview of the most recent developments on the experimental approach to the reaction.

---

[3] In addition to the campaigns discussed below there are also efforts underway to directly measure the reactions at the deep underground laboratories in China at the Jinping Underground laboratory (JUNA) and by CASPAR at the Sanford Underground Research Facility in the former Homestake mine in the USA.

Since the major nuclear-physics uncertainties are in whether there are low-energy resonances in the $^{22}$Ne($\alpha, \gamma$) $^{26}$Mg reaction, and in the branching/relative strength between the $\gamma$ and neutron channels for the $E_{c.m.}$ = 706 keV resonance, these are the major focus of the planned direct and indirect measurements.

## 4.1 EMMA-TIGRESS at TRIUMF

The EMMA experiment aims to use the $^{22}$Ne($^{7}$Li,$t$)$^{26}$Mg reaction in inverse kinematics to populate states in $^{22}$Ne($\alpha, \gamma$) $^{26}$Mg below the neutron threshold which may contribute to that reaction, such as the hypothetical $E_{c.m.}$ = 557 keV resonance from the study of Talwar et al. Stronger low-energy $^{22}$Ne($\alpha, \gamma$)$^{26}$Mg resonances will destroy some $^{22}$Ne which could be available for the $s$-process while also having some impact on magnesium isotopic ratios (see Ref. [48] and references therein). Past transfer reactions in normal kinematics [31,44,75] have been hindered by background resulting from the windows surrounding the gas cell or the substrate into which the neon target material was implanted. Measurements in inverse kinematics [10,33] have typically suffered from worse energy resolution due to target energy loss and angular aberrations in the reconstruction of the excitation energy. The EMMA-TIGRESS measurements will use the rejection power of the ElectroMagnetic Mass Analyser (EMMA) with the resolution of the high-purity germanium detectors of the TRIUMF-ISAC Gamma-Ray Escape-Suppressed Spectrometer (TIGRESS) to probe the low-energy $^{22}$Ne($\alpha, \gamma$)$^{26}$Mg resonances. In doing so, the background from windows/substrates and the resolution degradation from inverse kinematics can be overcome.

## 4.2 Texas A&M University Cyclotron Institute

Previous measurements at the Texas A&M University Cyclotron Institute include both an ANC measurement [33] and an transfer study which included a measurement of the heavy residues from the reaction [32]. Confirmation of these results, which resulted in a significant decrease in the neutrons from the $^{22}$Ne($\alpha, n$)$^{25}$Mg reaction, is highly desirable.

The planned measurement will follow on from Ref. [10] and use a very similar method to that experiment and also the EMMA-TIGRESS measurement mentioned above. The $^{22}$Ne($^{7}$Li,$t$)$^{26}$Mg reaction will be used to populate $^{26}$Mg states and the resulting $^{26}$Mg recoils will be transported through the MDM spectrometer and detected in the TexP-PACs detector at the focal plane. This experiment is focused on resolution and not yield. The silicon detectors have been placed further away from the target to try to limit the impact of angular aberration on the excitation-energy reconstruction, one of the limiting factors in the experiment reported in Ref. [10]. The targets are also being made to limit the inter-action region of the beam to improve the excitation-energy resolution and reduce the uncertainty in the branching ratio from the beam emittance.

**Fig. 7** Sketch of the particles involved in *Magn-a*, the first experimental attempt to get the low-energy $^{22}$Ne$(\alpha, n)^{25}$Mg cross section with THM. The experiment was performed at INFN-LNS [78]

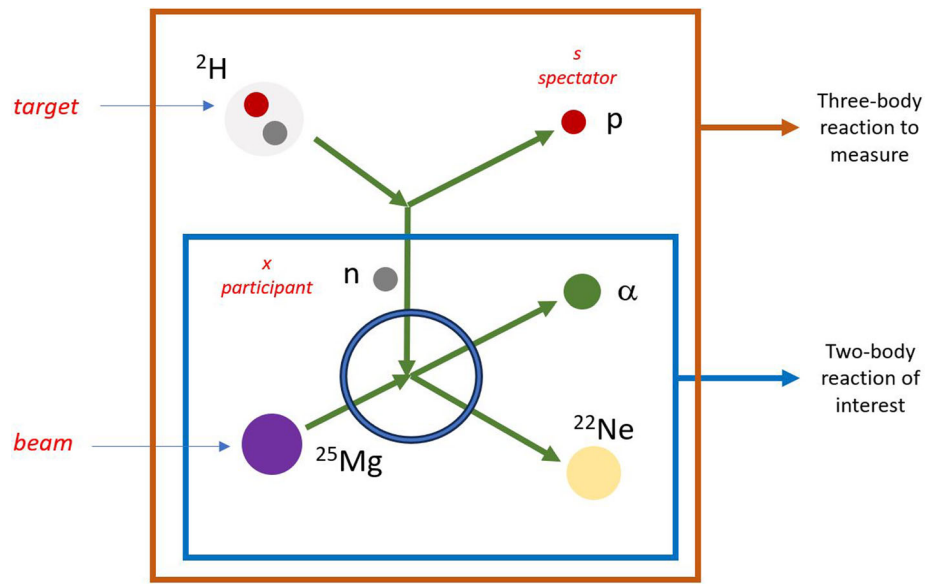

### 4.3 Trojan Horse measurements at the INFN-LNS

The Trojan Horse Method [76,77] offers an alternative pathway to measure the $S(E)$ factor down to the relevant astrophysical energies. To apply the method to $^{22}$Ne$(\alpha, n)^{25}$Mg, the AsFiN group at INFN-LNS plans to measure a two-to three-body reaction in quasi free kinematics, involving either $^{22}$Ne or $\alpha$, and the Trojan Horse (TH) nucleus. The TH nucleus exhibits a significant two-cluster configuration, where one cluster is either $\alpha$ or $^{22}$Ne ($x$, the participant cluster), while the other is spectator to the process, $s$.

A first attempt, the *Magn-a* experiment, was conducted at the INFN-LNS in Catania, Italy. This experiment aimed at measuring the $^2$H$(^{25}$Mg,p$\alpha)^{22}$Ne reaction to extract the $^{25}$Mg$(n, \alpha)^{22}$Ne cross section, and to retrieve the inverse reaction cross section via detailed balance. By using deuterons as virtual neutron sources, this approach circumvents the challenges associated with low neutron detection efficiency and degraded energy resolution typical of neutron beam experiments. In Fig. 7, the investigated process is illustrated, and experimental details of *Magn-a* can be found in [78].

Whilst data analysis is ongoing, other indirect measurements are planned, this time using implanted solid targets. Regarding the beam, the TH nuclei under investigation are $^{26}$Mg with the $^{22}$Ne+$\alpha$ ($x$+$s$) cluster configuration in $l = 0$ and $^{23}$Ne as $^{22}$Ne + p ($x$+$s$), provided that the experimental feasibility of $p$-wave intercluster motion is confirmed. It is intended to measure the TH cross section over the entire relevant astrophysical region ($E_{c.m.}$ from 1 MeV down to zero). Datasets from [19] will be used to normalize the data, as well as the future input from SHADES and other direct campaigns at low energies, which will strengthen the result in the ultra-low energy region of the excitation function.

### 4.4 Direct cross section measurements at the INFN-LNGS Bellotti Ion Beam Facility with the 3.5 MV Accelerator

A new campaign to directly measure the cross section of the 706 keV resonance and explore the lower energy region is currently underway at the Gran Sasso National laboratory of the INFN in the framework of the ERC starting grant SHADES. By measuring in the 4300 m.w.e. (meters-water-equivalent) underground location, the environmental neutron background is greatly reduced with respect to the flux on the Earth's surface [21], and the background rate is lower than the one of the most recent campaign by Jaeger et al. In addition, the newly inaugurated MV accelerator at the INFN Bellotti Ion Beam facility is able to provide high-intensity beams, in principle up to 500 $\mu$A, to further boost the signal-to-noise ratio [79]. A $^3$He-liquid scintillator[4] array surrounds a central extended, recirculating gas target containing > 99.995% enriched $^{22}$Ne. Eighteen $^3$He counters capture the reaction neutrons that were thermalised in twelve liquid scintillators, which also provide energy information if the neutrons are above the detection threshold. This is not expected to be the case for neutrons emitted from $^{22}$Ne$(\alpha, n)^{25}$Mg due to the negative $Q$ value but provides a handle on detecting if beam induced neutrons from above mentioned Be, C or B impurities are present.

---

[4] Scionix EJ-309 organic scintillation detector.

**Fig. 8** First results of the excitation function measurement of the $E_r = 706$ keV resonance at the Bellotti Ion Beam Facility at the LNGS. Note that the beam energy is currently under calibration and the displayed alpha energy is a nominal value

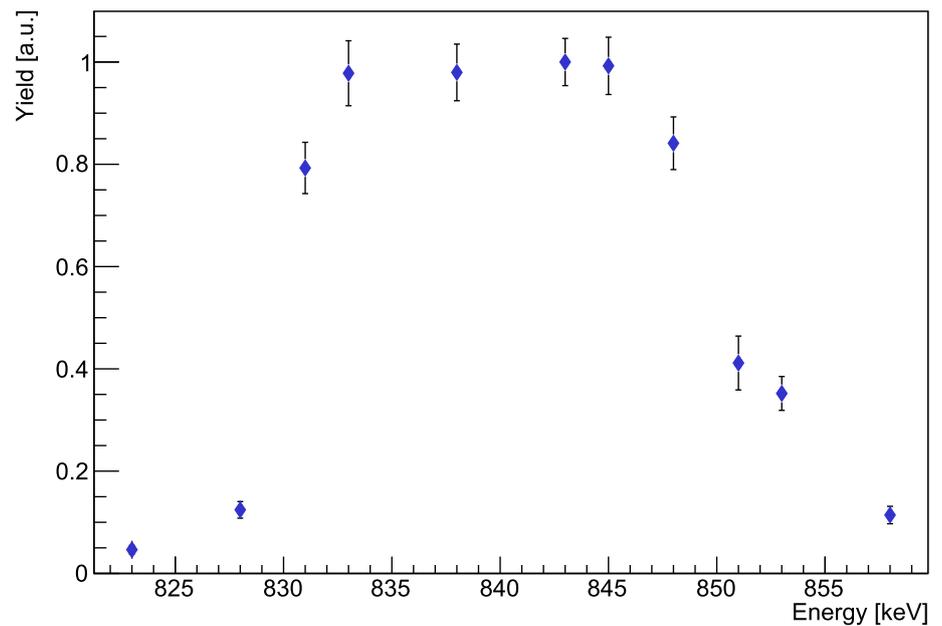

The goal of the currently ongoing campaign is the remeasurement of the 706 keV resonance (see Fig. 8) and the high-sensitivity probing of the low energy region, for example around the possible resonance at $E_r = 551$ keV. This campaign is running at the time of writing this manuscript, and the 706 keV resonance has so far been remeasured in detail, see Fig. 8. In particular, multiple resonance yield curves were measured using various target pressures resulting in target thicknesses, in terms of energy loss, between 1 keV and 30 keV. This will allow a precise resonance energy and strength determination as well as a new measurement of its total width. The efficiency calibration of the array is scheduled to take place at the ATOMKI laboratory in Debrecen, Hungary, using the activity method with the reaction $^{51}V(p,n)^{51}Cr$, as done previously for other neutron detector setups [30,80].

## 5 Conclusion

Despite the major effort devoted to investigating $^{22}Ne(\alpha, n)^{25}Mg$ and $^{22}Ne(\alpha, \gamma)^{26}Mg$ in the last decades there still remain unanswered questions about their reaction rates and astrophysical impact both for low-, intermediate mass and massive stars. More knowledge on the resonance parameters is required to reduce the uncertainty of the stellar rates to an acceptable level, as is pointed out in Sect. 3.

As has been discussed in Sect. 4, the problem is currently being approached from multiple angles, with a number of indirect and direct experiments underway or in planning stages. The hope is that through the concerted effort of mut-liple groups around the world the question of the existence of a resonance with significant strength at $E_r = 551$ keV can be answered and the strength and energy of the 706 keV resonance will converge on a high precision value.

**Funding** Open access funding provided by Università degli Studi di Napoli Federico II within the CRUI-CARE Agreement. The University of Naples "Federico II", the U.S. National Science Foundation project IReNA and ChETEC-INFRA (EU project no. 101008324) funded the meeting "The Big-Three Reactions for Astrophysics: 12C($\alpha$, $\gamma$)16O, 12C+12C fusion, 22Ne($\alpha$,n)25 Mg". RJD and MW were supported by the National Science Foundation through Grant No. PHY-2310059 (University of Notre Dame Nuclear Science Laboratory). AB, TC, DM and DR were supported through the European Union (ERC-StG SHADES #852016) and the Italian Ministry of Research (FARE project EAS$\gamma$ R20SLAA8CJ) - the latter also supported UB. This work was supported in part by the National Science Foundation under Grant No. OISE-1927130 (IReNA). SO acknowledges support from the Office of Nuclear Physics, Office of Science of the U.S. Department of Energy under Contract No. DE-AC02-98CH10886 with Brookhaven Science Associates, LLC. UB utilized resources provided by West-Grid (www.westgrid.ca) and Compute Canada Calcul Canada (www.computecanada.ca). The computing facilities (in particular the Orcinus cluster) was mostly supported by the UBC Chemistry Department Team led by Mark Thachuk. UB was supported by the European Union (ChETEC-INFRA, project no. 101008324).

**Code Availability Statement** This manuscript has no associated code/software. [Author's comment: There is no Code/Software associated with this study.]